\documentclass{appolb}

\usepackage{axodraw}
\newcommand{\nl}{\nonumber\\}

\newcommand{\lpar}{\left(}                            
\newcommand{\rpar}{\right)}

\newcommand{\bq}{\begin{equation}}                    
\newcommand{\eq}{\end{equation}}
\newcommand{\bqa}{\arraycolsep 0.14em\begin{eqnarray}}
\newcommand{\eqa}{\end{eqnarray}}
\newcommand{\ba}[1]{\begin{array}{#1}}
\newcommand{\ea}{\end{array}}
\newcommand{\ben}{\begin{enumerate}}
\newcommand{\een}{\end{enumerate}}
\newcommand{\bei}{\begin{itemize}}
\newcommand{\eei}{\end{itemize}}
\newcommand{\eqn}[1]{Eq.(\ref{#1})}

\newcommand{\fig}[1]{Fig.~\ref{#1}}

\def\Re{\mathop{\operator@font Re}\nolimits}
\def\Im{\mathop{\operator@font Im}\nolimits}

\newcommand{\spro}[2]{{#1}\cdot{#2}}
\newcommand{\egam}[1]{\Gamma\lpar#1\rpar}               

\newcommand{\ep}{\epsilon}

\newcommand{\Reb}{{\rm{Re}}}

\newcommand{\upar}[1]{u}

\newcommand{\ssN}{{\scriptscriptstyle{N}}}

\newcommand{\bqas}{\begin{eqnarray*}}
\newcommand{\eqas}{\end{eqnarray*}}

\def\app#1#2 {{\it Acta. Phys. Pol.} {\bf#1},#2}
\def\cpc#1#2 {{\it Computer Phys. Comm.} {\bf#1},#2}
\def\np#1#2 {{\it Nucl. Phys.} {\bf#1},#2}
\def\pl#1#2 {{\it Phys. Lett.} {\bf#1},#2}
\def\prep#1#2 {{\it Phys. Rep.} {\bf#1},#2}
\def\prev#1#2 {{\it Phys. Rev.} {\bf#1},#2}
\def\prl#1#2 {{\it Phys. Rev. Lett.} {\bf#1},#2}
\def\zp#1#2 {{\it Zeit. Phys.} {\bf#1},#2}
\def\sptp#1#2 {{\it Suppl. Prog. Theor. Phys.} {\bf#1},#2}
\def\mpl#1#2 {{\it Modern Phys. Lett.} {\bf#1},#2}
\def\jetp#1#2 {{\it Sov. Phys. JETP} {\bf#1},#2}
\def\fpj#1#2 {{\it Fortschr. Phys.} {\bf#1},#2}
\def\afp#1#2 {{\it Acta.Phys. Polon.} {\bf#1},#2}
\def\err#1#2 {{\it Erratum} {\bf#1},#2}
\def\ijmp#1#2 {{\it Int. J. Mod. Phys} {\bf#1},#2}
\def\nc#1#2 {{\it Nuovo Cimento} {\bf#1},#2}
\def\ap#1#2 {{\it Ann. Phys.} {\bf#1},#2}
\def\cmp#1#2 {{\it Comm. Math. Phys.} {\bf#1},#2}
\def\el#1#2 {{\it Europhys. Lett.} {\bf#1},#2}
\def\hpa#1#2 {{\it Helv. Phys. Acta} {\bf#1},#2}
\def\yf#1#2 {{\it Yad. Fiz.} {\bf#1},#2}
\def\nim#1#2 {{\it Nucl. Instrum. Meth.} {\bf#1},#2}
\def\spz#1#2 {{\it Sov. Pisma Zhetf} {\bf#1},#2}
\def\jetpl#1#2 {{\it JETP Lett.} {\bf#1},#2}
\def\sjnp#1#2 {{\it Sov. J. Nucl. Phys.} {\bf#1},#2}
\def\ptp#1#2 {{\it Progr. Theor. Phys. (Kyoto)} {\bf#1},#2}
\def\rmp#1#2  {{\it Rev. Mod. Phys.} {\bf#1},#2}
\def\zhetf#1#2 {{\it ZhETF} {\bf#1},#2}
\def\prs#1#2 {{\it Proc. Roy. Soc.} {\bf#1},#2}
\def\phys#1#2 {{\it Physica} {\bf#1},#2}

\def\bfi{\begin{figure}}
\def\efi{\end{figure}}

\newcommand{\hyper}[4]{{}_2F_1(#1\,,\,#2\,;\,#3\,;\,#4)}

\newcommand{\intsx}[1]{\int_{\scriptstyle 0}^{\scriptstyle 1}\!\!\!d#1}
\newcommand{\intsxy}[2]{\int_{\scriptstyle 0}^{\scriptstyle 1}\!\!\!d#1
                        \int_{\scriptstyle 0}^{\scriptstyle #1}\!\!\!d#2}

\begin{document}
\pagestyle{plain}
\newcount\eLiNe\eLiNe=\inputlineno\advance\eLiNe by -1
\title{Numerical evaluation of non-infrared two-loop vertices%
\thanks{
Talk given at the final meeting of the European Network 
``Physics at Colliders'', Montpellier, September 26-27, 2004.
This work has been supported by the European Community's Human 
Potential Programme under contract HTRN-CT-2000-00149 Physics at 
colliders.}%
}
\author{Sandro Uccirati\footnote{uccirati@mppmu.mpg.de}
\address{
Max-Planck-Institut f\"ur Physik 
(Werner-Heisenberg-Institut),\\
F\"ohringer Ring 6,
80805 M\"unchen
}}
\maketitle

\begin{abstract}
Some methods for the numerical computation of two-loop non-infrared vertices 
are reviewed.
A new method is also proposed and compared to the old ones.
Finally, some preliminary results are presented, concerning the evaluation 
of the fermionic corrections to $\sin^2\theta_{\rm eff}^{\rm lept}$ 
throught the described techniques.
\end{abstract}

\section{Introduction}
In the forthcoming experiments, the validity of the Standard Model 
will be tested with high precision.
In addition to the direct search of the Higgs Boson, important 
quantities will be measured in the future colliders, providing a good 
test of the Model.
This of course pushes the theorists to compute these observables at 
the same degree of precision.
For example, the mass of the W boson, whose present value is 
$M_W=80.426\pm 0.034$ GeV (\cite{unknown:2003ih}), will be measured with an 
expected error of $15$ MeV at LHC and $6$ MeV at the ILC.
Or the effective leptonic weak mixing angle 
($\sin^2\theta_{\rm eff}^{\rm lept}= 0.23150(16),$\cite{unknown:2003ih}) 
will be known with an absolute precision of $10^{-5}$ at the ILC.
To get a similar theoratical incertenty, we have to improve the 
calculation in perturbation theory beyond the one-loop level.
The computation of two-loop Feynman diagrams is a hard task.
The pure analytical techniques are very efficient when few mass scales are 
present (see for example~\cite{Davydychev:2002} or~\cite{Remiddi:2003}), 
but seem to be unable to deal with the complete set 
of two-loop diagrams in the Standard Model (where more scales come 
into the game).
For this reason we were led to abandon the analytical way in favor of  
a numerical evaluation of multi-loop diagrams.
The goal of the numerical approach is to express any diagram in terms 
of smooth integrals.
\section{Standard BT relation\label{oldBT}}
The Bernstein-Tkachov theorem~\cite{Tkachov:1997wh} tells us that for any 
finite set of polynomials $V_i(x)$, where
$x = \,\lpar x_1,\dots, x_{\ssN}\rpar$ is a vector of Feynman parameters,
there exists an identity of the following type:
\bq
{\cal F}\,\lpar x,\partial\rpar \prod_i\,V_i^{\mu_i+1}(x) =
B\,\prod_i\,V_i^{\mu_i}(x),
\label{functr}
\eq
where ${\cal F}$ is a polynomial of $x$ and $\partial_{i} = \partial/\partial
x_i$; $B$ and all coefficients of ${\cal F}$ are polynomials of $\mu_i$ and of
the coefficients of $V_i(x)$.
If the polynomial $V$ is of second degree we have a master formula,
again due to F.~V.~Tkachov~\cite{Tkachov:1997wh}.
We write the quadratic $V$ as:
\bq V(x) = x^t\,H\,x + 2\,K^t\,x + L,
\label{defHKL}
\eq
where $x^t=(x_1,...,x_n)$, $H$ is an $n \times n$ matrix, $K$ is an $n$
vector.
The solution to the problem of determining the polynomial ${\cal F}$ is
as follows:
\bq
{\cal F}=
1 + \frac{{\cal P}^t\,\partial_x}{\mu+1},
\qquad\qquad
{\cal P}= -\,\frac{x-X}{2},
\label{P}
\eq
\bq
B = L - K^t\,H^{-1}\,K
\qquad\qquad
X= - H^{-1}\,K.
\label{BX}
\eq
Therefore we have:
\bq
V^{\mu}(x)= 
\frac{1}{B}\,
\Big[
1 + \frac{{\cal P}^t\,\partial_x}{\mu+1}
\Big]
V^{\mu+1}(x).
\label{BT}
\eq
This is the standard BT relation for quadratics.
\section{Strategy}
The standard BT relation is very usefull for the computation of one-loop 
diagrams and also some two-loop configurations 
(see~\cite{Passarino:2001wv}, \cite{Passarino:2001jd}, 
\cite{Ferroglia:2002mz}, \cite{Ferroglia:2002yr}, \cite{Ferroglia:2003yj}).
The strategy is the following.
First of all, the diagram with $N$ external legs and a tensorial 
structure of rank $n$ is decomposed in form factors:
\bq
G_{\ssN}^{\mu_1,\dots,\mu_n}= \sum_{T}\,G_{\ssN}(T)\,T^{\mu_1,\dots,\mu_n}.
\eq
Here $T^{\mu_1,\dots,\mu_n}$ are all possible tensors of rank $n$ 
that can be obtained by combining the $N-1$ independent external momenta 
of the diagram and the metric $\delta_{\mu\,\nu}$.

For one-loop diagrams, if we write each form factor in the parametric 
space, we always obtain a result of this form (see~\cite{Ferroglia:2002mz} 
for details):
\bq
G_{\ssN}=
\left( \frac{\mu^2}{\pi} \right)^{\ep/2}\!\!\!\!
\egam{n+\frac{\ep}{2}}\!
\int\!dS_{\ssN-1}(x)\,
P(x)\,V(x)^{-n-\ep/2}
\qquad
n \in I\!\!N,
\label{FF1L}
\eq
where
\bq
\int\!dS_m(z) = \intsxy{z_1}{x_2}\,...\,\int_0^{z_{m-1}}dz_m,
\eq
$\mu$ is the mass scale and $P$ and $V$ are polinomials 
in the Feynman parameters $x=(x_1,\dots,x_{\ssN-1})$.
In particular $V(x)$ is always a quadratic of the type \eqn{defHKL}.
The $\Gamma$ function contains the UV pole (if present) of the diagram.
Its argument is always equal to the exponent of $V$ with opposite sign 
and therefore, for every UV divergent one-loop diagram, $V$ has a vanishing 
power.

The goal is to express $G_{\ssN}$ in terms of smooth integrals to be 
integrated numerically. For $n=0$ (which corresponds to UV divergent 
form factors), we can simply perform a Laurent expansion around $\ep=0$ 
to get just smooth integrands of the type:
\bq
P(x)\,\ln^k V(x).
\eq
For $n \ge 1$ the idea is to apply the BT relations \eqn{BT} to ``raise'' 
the power of $V$ of one unit and then integrate by parts to get rid of the 
derivatives.
Then the procedure is repeated for all integrals that are generated, until 
the power of $V$ becomes $-\ep/2$, and, at this point, we proceed as in the 
case $n=0$.

The procedure is clear and leads to smooth integrals at the price of 
introducing the denominator $B$ which of course can vanish somewhere in 
the phase space.
It can be prooved that the zeros of $B$ correspond to the leading Landau 
singularity of the diagram, but the singular behaviour is usually 
overestimated by the BT procedure (see~\cite{Ferroglia:2002mz}). 
This means that in the region $B\sim0$ all terms generated by \eqn{BT} 
cancel one another, giving rise to numerical instabilities.
For this reason it would be good to find a new relation, which should 
contain the real divergent behaviour for $B=0$ and therefore should 
remain stable also for $B\sim0$.

In the two-loop case the form factors are classified counting the 
number of propagator of each loop:
\bq
G_{\ssN}(T) \quad\to\quad G_{a\,b\,c}(T),
\eq
where $a$ is the number of propagators (with momentum $q_1$) which belong just 
to the first loop (the one with the smallest number of propagators), 
$b$ is the number of propagators (with momentum $q_2$) which belong just 
to the second loop and 
$c$ is the number of propagators (with momentum $q_1-q_2$) which belong 
to both loops.

Then we first parametrise the loop with momentum $q_1$, obtaining 
a new propagator in $q_2$ with a non integer power. 
The mass and the momentum of this new propagator depend in general on 
the Feynam parameters of the first loop. 
After the parametrisation of the second loop, each form factor takes the 
form ( see~\cite{Actis:2004bp} for details):
\bqa
G_{a\,b\,c}
&=&
{\cal A}_{\ep}\,\left( \frac{\mu^2}{\pi} \right)^{\ep}\,\egam{n+\ep}
\int\!dS_{a+c-1}(x)\,\int\!dS_{b}(x)\,
\nl
&{}&
[x_2\,(1-x_2)]^{-m-\ep/2}\,y_b^{m-1+\ep/2}
P(x,y)\,V_x(y)^{-n-\ep},
\label{Gacb}
\eqa
where $m,n \in I\!\!N$, $\mu$ is again the mass scale and 
${\cal A}_{\ep}$ is a constant regular in $\ep=0$.
$P$ is a polinomial in all Feynman parameters $x=(x_1,\dots,x_{a+c-1})$ 
and $y=(y_1,\dots,y_b)$.
$V_x(y)$ is a quadratic of the type \eqn{defHKL} in $y$, where now the 
coefficients $H$, $K$, $L$ depend on $x$ and have the following form:
\bq
\frac{C(x)}{x_2^h\,(1-x_2)^k},
\eq
where $h,k \in I\!\!N$ and $C$ is quadratic in $x$.
The $\Gamma$ function contains the overall UV pole (if present),
while sub-divergencies are contained in 
$[x_2\,(1-x_2)]^{-m-\ep/2}\,y_b^{m-1+\ep/2}$.
Since at least one of the coefficients of $V_x$ have in the denominator the 
product $x_2\,(1-x_2)$, the divergent behaviuor of the integrand at $x_2=0$ 
and $x_2=1$ is present (generating the UV pole) just for $m>n$.
On the other hand the UV divergency is generated by the behaviour in $y_b=0$
only for $m=0$.

So, apart the UV pole coming from sub-loops, any two-loop diagram is 
the integral of a one-loop diagram whose masses and momenta depend on 
the integration variables.
If we would be able to express any one-loop diagram in terms of smooth 
integrals not only with respect to the integration variables, but 
also with respect to their external masses and momenta, the numerical 
evaluation of two-loop diagrams would be a trivial task.
The BT relation \eqn{BT} is in general not good for this purpose, 
because it introduces the denominator $B$.
In fact, since the coefficients $H$, $K$ and $L$ of $V_x$ (\eqn{Gacb}) 
depend on the Feynman variables $x$, the same happens for $B$ which 
therefore generates singularities inside the $x$ integration contour.
As a consequence, apart some special cases where it is possible to have 
a factor $B$ independent of any Feynman variable, the standard BT method can 
not be applied for two-loop diagrams.
This is another reason to search for a new BT-like relation.
\section{The new BT-like relation\label{newBT}}
To obtain new relations it is usefull to write the quadratic $V(x)$ 
in the following way:
\bqa
V(x) 
&=&
x^t\,H\,x + 2\,K^t\,x + L 
\nl
&=&
(x^t-X^t)\,H\,(x-X) + B= 
Q(x) + B.
\label{defQ}
\eqa
This formula, which defines the quadratic $Q(x)$, can be trivially verified 
using the definition of $B$ and $X$ in \eqn{BX}.
The basic relation satisfied by $Q(x)$ is the following:
\bq
{\cal P}^t\,\partial_x\,Q(x)= - Q(x),
\qquad\qquad
{\cal P}= -\,\frac{x-X}{2}.
\label{BTQ}
\eq
At this point we introduce a new variable $y$ and a new polynomial 
$W(x,y)$ defined as follows:
\bq
W(x,y)= Q(x)\,y + B,
\eq
and satisfing the following relation:
\bq
( {\cal P}^t\,\partial_x + y\,\partial_y )\,W^{\mu}(x,y)= 0.
\label{BTW}
\eq
Next we consider the following integral
\bq
I_{\beta}= \intsx{y}\,y^{\beta-1}\,W^{\mu}(x,y),
\qquad\qquad
\beta > 0
\eq
and compute:
\bqa
\!\!\!\!\!\!\!\!\!\!\!\!
{\cal P}^t\,\partial_x\,I_{\beta} 
&=&
\intsx{y}\,y^{\beta-1}\,{\cal P}^t\,\partial_x\,W^{\mu}(x,y)=
-\intsx{y}\,y^{\beta}\,\partial_y\,W^{\mu}(x,y)
\nl
&=&
- W^{\mu}(x,1) + \beta\intsx{y}\,y^{\beta-1}\,W^{\mu}(x,y)= 
- V^{\mu}(x) + \beta\,I_{\beta}.
\eqa
Using the definition of the hypergeometric function 
(see~\cite{ellip}) to evaluate $I_{\beta}$, we finally get:
\bq
V^{\mu}(x)= 
B^{\mu}\,\left( 1 - \frac{{\cal P}^t\,\partial_x}{\beta}\, \right)
\hyper{-\mu}{\beta}{\beta+1}{-\frac{Q}{B}}.
\label{BTnew}
\eq
This formula has a general validity and does not depend to the 
actual expression for $Q$, $B$ and ${\cal P}$. 
The only relations which they must satisfy are:
\bq
V(x)= Q(x) + B,
\qquad\qquad
{\cal P}^t\,\partial_x\,Q(x)= - Q(x),
\qquad\qquad
\beta > 0.
\eq
The usefullness of this relation is evident if we consider the 
case $\mu= - 1 - \alpha\,\ep$.
In this case the better choice for the free parameter $\beta$ is 
$\beta=1$. 
Using the expansion of the hypergeometric function around
$\ep=0$
\bq
\hyper{1+\alpha\,\ep}{1}{2}{x} =
- \frac{1}{x}\,\sum_{n=0}^{\infty}\,
\frac{(-\alpha\,\ep)^n}{(n+1)!}\,\ln^{n+1}(1-x),
\eq
we obtain
\bq
V^{-1-\alpha\,\ep} = 
\sum_{n=1}^{\infty}\,\frac{(-\alpha\,\ep)^{n-1}}{n!}\,
\left(1 - {\cal P}^t\,\partial_x \right)\,
\frac{B^{-\alpha\,\ep}}{Q}\,\ln^n\left( 1 + \frac{Q}{B} \right).
\label{BTnew,mu=-1}
\eq
In this relation we have obtained our goal to avoid the appearence of the 
factor $B$ in the denominator. 
Here, the only denominator is $Q(x)$ which can vanish inside the 
integration contour;
however its zeros are compensated by the logarithm, whose argument goes to 
1 when $Q(x)$ goes to 0.

An example of the usefullness of the new relation is the evaluation of 
one-loop three-point functions.
In the scalar case we have:
\bq
C_0=
\left( \frac{\mu^2}{\pi} \right)^{\ep/2}\!\!
\egam{1+\frac{\ep}{2}}\!
\intsxy{x_1}{x_2}\,V(x_1,x_2)^{-1-\ep/2}.
\eq
If we insert \eqn{BTnew,mu=-1} and integrate by parts, we simply obtain:
\bq
C_0= 
\sum_{i=0}^{2}\,\frac{a_i}{2}\,\intsx{x}\,\frac{1}{Q[i](x)}\,
\ln\left( 1 + \frac{Q[i](x)}{B} \right)
+ {\cal O}(\ep),
\label{C0}
\eq
where
\bq
Q[0](x)= Q(1,x),
\qquad
Q[1](x)= Q(x,x),
\qquad
Q[2](x)= Q(x,0),
\eq
\bq
a_0= 1-X_1,
\qquad
a_1= X_1-X_2,
\qquad
a_2= X_2.
\eq
This result for $C_0$ (which can be easily generalised for tensor 
integrals) can be also used to compute two-loop diagrams which can be 
expressed as an integral of a one-loop three-point function.
We see for example what happens in two families of two-loop vertices.
\section{The two-loop vertex $V^{131}$}
All two-loop vertices can be classified according to six families. 
Their list is given in the appendix.
Taking into consideration the $V^{131}$ vertex, after Feynman 
parametrisation it takes the form:
\bq
V^{131} =
- \left( \frac{\mu^2}{\pi} \right)^{\ep}\!\!\egam{1+\ep}\!
\int_0^1\!\!\!dx\!\int\!\!dS_3(y,z_1,z_2)\,
[x(1-x)]^{-\ep/2}(1-y)^{\ep/2-1}\,U^{-1-\ep},
\eq
\bq
U = (z^t-X^t)\,H\,(z-X) + (m_x^2-m_3^2)\,(1-y) + B,
\qquad
m_x^2 = \frac{m^2_1}{x} + \frac{m^2_2}{1-x},
\eq
where the masses are defined in the figure for $V^{131}$ in the appendix. 
The coefficient $H$, $X$ and $B$ are those appearing in the polynomial 
of a one-loop three-point funcion, with external momenta $-P$, $p_1$, $p_2$ 
and masses $m_3$, $m_4$, $m_5$.
If we introduce $Q(y,z_1,z_2)$ for the polynomial $U$ defined by:
\bq
U(y,z_1,z_2)= Q(y,z_1,z_2) + B,
\eq
we see that $Q$ satisfies the following basic relation:
\bq
\Big[
(1-y)\,\partial_y - \frac{(z^t-X^t)\,\partial_z}{2}
\Big]
\,Q=
-Q.
\eq
From this formula we obtain the standard BT relation and the new one 
(choosing $\beta=1$):
\bq
U^{-1-\ep} = 
\frac{1}{B}\,\left[
1 
- \frac{(1-y)\,\partial_y}{\ep}
+ \frac{(z^t-X^t)\,\partial_z}{2\,\ep}
\right]\,
U^{-\ep},
\label{v131bt}
\eq
\bq
U^{-1-\ep} =
\sum_{n=1}^{\infty}\,\frac{(-\ep)^{n-1}}{n!}\,
\left[
1 - (1-y)\,\partial_y + \frac{(z^t-X^t)\,\partial_z}{2}
\right]\,
\frac{B^{-\ep}}{Q}\,
\ln^n\left( 1 + \frac{Q}{B} \right).
\label{v131btnew}
\eq
From these equations, we see that $V^{131}$ is exactly one of those 
particular two-loop configurations for which can be found a factor 
$B$ independent from any Feynman parameter.
If this is crucial for the application of the standard BT method 
(\eqn{v131bt}), this would not be strictly required for the new 
method (\eqn{v131btnew}).
In addition to that the new relation appear to have a better behaviour 
near the zeros of $B$.

However, for this type of diagrams, where the polynomial $U$ is linear 
in one of the Feynman variables ($y$), another procedure is available.
After the transformation $y \to 1-(1-z_1)\,y$, we have to compute:
\bq
I=
\intsx{y}\,y^{\ep/2-1}\,(a\,y+b)^{-1-\ep}= 
\frac{2\,b^{-1-\ep}}{\ep}\,
\hyper{1+\ep}{\frac{\ep}{2}}{\frac{\ep}{2}+1}{-\frac{a}{b}},
\eq
\bq
a = (m_x^2-m_3^2)\,(1-z_1),
\qquad
b = (z^t-X^t)\,H\,(z-X) + B.
\eq
By applying the properties of the hypergeometric function 
(see~\cite{Ferroglia:2003yj} for details) and expanding in $\ep$ we 
simply obtain:
\bqa
\!\!\!\!\!\!\!\!
V^{131}
&=&
\intsxy{z_1}{z_2}\,\frac{1}{V}\,
\ln\left( 1 + \frac{V}{(m_x^2-m_3^2)\,(1-z_1)} \right)
\nl
&{}&
+ \left( -\frac{2}{\ep} + \intsx{x}\,\ln A_x \right)\,
C_0(P^2,p_1^2,p_2^2,m_3,m_4,m_5),
\label{v131hy}
\eqa
where
\bq
V= (z^t-X^t)\,H\,(z-X) + B,
\qquad
A_x= - m_3^2\,x\,(1-x) + m_1^2\,(1-x) + m_2^2\,x.
\eq
Note that $V$ is exactly the polynomial of the $C_0$ function 
appearing in \eqn{v131hy}.
The numerical results for the three methods are then compared 
in \fig{tablev131}.
\begin{figure}[ht]
\begin{center}
\begin{tabular}{|c|c|c|c|c|c|c|c|l|}
\hline &&&&&&&& \\
$s$ & $p_1^2$ & $p_2^2$ & 
$m_1$ & $m_2$ & $m_3$ & $m_4$ & $m_5$ &
$\Reb\,V^{131}_0$ \\
       &&&&&&&& \\
\hline
\hline &&&&&&&& \\
$100^2$ & $m_b^2$ & $m_b^2$ & 
$m_{_W}$ & $m_{_W}$ & $m_{_Z}$ & $m_b$ & $m_{_Z}$ 
       & $0.274717(2)\times 10^{-2}$ \\
&&&&&&&& $0.2747182(5)\times 10^{-2}$ \\
&&&&&&&& $0.2747194\times 10^{-2}$ \\
\hline &&&&&&&& \\
$800^2$ & $m_b^2$ & $m_b^2$ & 
$m_{_W}$ & $m_{_W}$ & $m_{_Z}$ & $m_b$ & $m_{_Z}$ 
       & $-0.247(4)\times 10^{-3}$ \\
&&&&&&&& $-0.2456(9)\times 10^{-3}$ \\
&&&&&&&& $-0.24612(8)\times 10^{-3}$ \\
\hline
\hline &&&&&&&& \\
$500^2$ & $m_t^2$ & $m_t^2$ & 
$m_{_W}$ & $m_{_W}$ & $m_{_Z}$ & $m_t$ & $m_{_Z}$ 
       & $0.952(1)\times 10^{-5}$ \\
&&&&&&&& $0.9536(7)\times 10^{-5}$ \\
&&&&&&&& $0.9545(13)\times 10^{-5}$ \\
\hline
\hline &&&&&&&& \\
$m_{_Z}^2$ & $m_e^2$ & $m_e^2$ & 
$m_t$ & $m_t$ & $m_{_Z}$ & $m_e$ & $m_{_Z}$ 
       & $0.288416(3)\times 10^{-2}$ \\
&&&&&&&& $0.288418(3)\times 10^{-2}$ \\
&&&&&&&& $0.2884222\times 10^{-2}$ \\
\hline
\hline &&&&&&&& \\
$m_{_Z}^2$ & $m_e^2$ & $m_e^2$ & 
$m_t$ & $m_t$ & $0$ & $m_e$ & $0$ 
       & $unstable$ \\
&&&&&&&& $-0.195(47)$ \\
&&&&&&&& $-0.214261(7)$ \\
\hline
\hline &&&&&&&& \\
$m_{_Z}^2$ & $m_e^2$ & $m_e^2$ & 
$m_e$ & $m_e$ & $0$ & $m_e$ & $0$ 
       & $unstable$ \\
&&&&&&&& $0.2080(3)$ \\
&&&&&&&& $unstable$ \\
\hline
\end{tabular}
\end{center}
\caption[]{Numerical results for the $V^{131}$ family: 
$V^{131}=\, V^{131}_{-1}\,\ep^{-1} + V^{131}_0$. All momenta are given 
in GeV. The first entry refers to the standard BT method \eqn{v131bt}, 
the second to the new method \eqn{v131btnew} and the third to 
\eqn{v131hy}.
When no number appears in curly brackets, this means that the error does not 
affect the written digits.
In the last two cases we have $B\sim0$ and in the last case also 
$m_x^2-m_3^2\sim0$.}
\label{tablev131}
\end{figure}
\section{The two-loop vertex $V^{231}$}
Another imprtant example of two-loop vertex topologies is $V^{231}$ 
(see appendix).
In this case Feynman parametrisation gives:
\bqa
V^{231} 
&=&
- \left( \frac{\mu^2}{\pi} \right)^{\ep}\!\!\egam{2+\ep}\!
\int\!dS_2(x_1,x_2)\,[x_2(1-x_2)]^{-1-\ep/2}
\nl
&{}&
\int\!\!dS_3(y_1,y_2,y_3)\,y_3^{\ep/2}\,U^{-2-\ep},
\eqa
where
\bqa
U 
&=&
- [ p_2\,y_1 - P\,(y_2-{\cal X}\,y_3) + p_1 ]^2
+ ( P^2 - p_1^2 + m_6^2 - m_5^2 )\,y_1
\nl
&{}&
- ( P^2 + m_6^2 - m_4^2 )\,y_2
+ ( M_x^2 - m_4^2 )\,y_3
+ p_1^2 + m_5^2,
\eqa
\bq
{\cal X}= \frac{1-x_1}{1-x_2},
\quad
M^2_x = \frac{- P^2\,x^2_1 + x_1\,( P^2 + m^2_1 - m^2_2) +
x_2\,( m^2_3 - m^2_1) + m^2_2}{x_2\,(1 - x_2)}.
\eq
Now the polynomial $U$ is a quadratic in $y_1$, $y_2$ and $y_3$ of the type 
$U= y^t\,H\,y + 2\,K^t\,y + L$ and in principle we could apply the 
new BT-like relation for $\mu=-2-\ep$.
Nevertheless, this would not be the most clever approach, because in this
case the $H$ matrix is singular and therefore the BT-factrors $B$ and 
$X$ are not well-defined.
Anyway, thanks to this singularity, we can perform the following change of 
variable
\bq
y_2 \to y_2 + {\cal X}\,y_3
\eq
and obtain a new polynomial $U'$ which is now linear in $y_3$.
Since this diagram is not UV divergent, we can set $\ep=0$ and the 
integration in $y_3$ becomes trivial, giving:
\bqa
V^{231}
&=&
\int\!dS_2(x_1,x_2)\int_0^1\!\!dy_1\,\frac{1}{A(x)}\,
\\
&{}&
\left\{
  \int_{(1-{\cal X})y_1}^{y_1}\!\!\!\!dy_2\!
  \left[ \frac{x_2\,(1-x_1)}{(y_1-y_2)\,A(x) + x_2\,(1-x_1)\,B(y)}
       - \frac{1}{B(y)} \right]
\right.
\\
&{}&
\left.
+ \int_0^{(1-{\cal X})y_1}\!\!\!\!dy_2\!
  \left[ \frac{x_2\,(x_1-x_2)}{y_2\,A(x) + x_2\,(x_1-x_2)\,B(y)}
       - \frac{1}{B(y)} \right]
\right\},
\eqa
where $A(x)$ and $B(y)$ are quadratics rispectively in $x_1$, $x_2$ and 
$y_1$, $y_2$.
Therefore each term is now a quadratic in $y_1$, $y_2$ to power $-1$ 
with $x$-dependent coefficients.
In other words, the $y$-integrations are one-loop 3-point functions $C_0$ 
with $x$-dependent masses and momenta (of course some change of variable 
has to be done to reduce to the usual simplex in $y_1$, $y_2$).
Note that the zeros of $A(x)$ do not spoil the smoothness of the integrand 
because the C functions cancel one another in the limit $A(x)\to0$.
At the end the diagram is written in the following form
\bq
V^{231}= 
\int\!dS_2(x_1,x_2)\,\frac{1}{A(x)}\,
\Big[
C_0^{(1)} - C_0^{(2)} + C_0^{(3)} - C_0^{(4)}
\Big]
\eq
and computed numerically using for the $C_0$ functions the expression
of \eqn{C0} obtained with the new method.
The numerical results are shown in \fig{tablev231}.
\begin{figure}[ht]
\begin{center}
\begin{tabular}{|c|c|l|}
\hline && \\
$s$, $-p_1^2$, $-p_2^2$ &
$m_1$, $m_2$, $m_3$, $m_4$,$m_5$, $m_6$
& $\Reb\,V^{231}$ \\
       && \\
\hline
\hline && \\
$100^2$, $m_b^2$, $m_b^2$ & 
$m_b$, $m_b$, $m_{_Z}$, $m_b$, $m_{_Z}$, $m_b$ 
 & $0.5028(4) \times 10^{-7}$ \\
\hline && \\
$200^2$, $m_b^2$, $m_b^2$ &
$m_b$, $m_b$, $m_{_Z}$, $m_b$, $m_{_Z}$, $m_b$ 
 & $0.9816(20) \times 10^{-8}$ \\
\hline
\hline && \\
$8m_t^2$, $m_t^2$, $m_t^2$ & 
$m_t$, $m_t$, $m_{_Z}$, $m_t$, $m_{_Z}$, $m_t$ 
 & $0.1489(5) \times 10^{-8}$ \\
\hline && \\
$20m_t^2$, $m_t^2$, $m_t^2$ & 
$m_t$, $m_t$, $m_{_Z}$, $m_t$, $m_{_Z}$, $m_t$ 
 & $0.1675(6) \times 10^{-8}$ \\
\hline
\hline && \\
$m_{_Z}^2$, $m_e^2$, $m_e^2$ & 
$m_t$, $m_t$, $m_t$, $m_{_Z}$, $m_e$, $m_{_Z}$ 
 & $-0.2018966 \times 10^{-8}$ \\
\hline
\hline && \\
$m_{_Z}^2$, $m_e^2$, $m_e^2$ & 
$m_t$, $m_t$, $m_t$, $0$, $m_e$, $0$ 
 & $0.5987(6)\times 10^{-6}$ \\
\hline
\hline && \\
$m_{_Z}^2$, $m_e^2$, $m_e^2$ & 
$m_e$, $m_e$, $m_e$, $0$, $m_e$, $0$ 
 & $-0.21161(49) \times 10^{-3}$ \\
\hline
\end{tabular}
\end{center}
\caption[]{Numerical results for the $V^{231}$ family. 
All momenta are given in GeV.
When no number appears in curly brackets, this means that the error 
does not affect the written digits.}
\label{tablev231}
\end{figure}
\section{Work in progress: fermionic correction to 
$\sin^2\theta_{\rm eff}^{\rm lept}$}
\begin{figure}[ht]
{\small
\noindent
\begin{tabular}{ccc}
\begin{picture}(100,50)(0,0)
\Photon(0,0)(30,0){1.3}{6}
\Line(90,34.5)(30,0)
\Line(90,-34.5)(30,0)
\Photon(70,10)(70,23){1.3}{3}
\Photon(70,-23)(70,-10){1.3}{3}
\CArc(70,0)(10,0,360)
\Text(5,-12)[cb]{$\mu$}
\Text(86,14)[cb]{${\scriptstyle \gamma,Z,W}$}
\Text(86,-18)[cb]{${\scriptstyle \gamma,Z,W}$}
\Text(15,30)[cb]{$a$}
\end{picture}
&\quad
\begin{picture}(100,50)(0,0)
\Photon(0,0)(30,0){1.3}{6}
\Line(90,34.5)(70,23)
\Photon(70,23)(58.5,16.3){1.3}{3}
\Photon(41,7)(30,0){1.3}{3}
\CArc(50,11.5)(10,0,360)
\Line(90,-34.5)(70,-23)
\Photon(70,-23)(30,0){1.3}{8}
\Line(70,-23)(70,23)
\Text(5,-12)[cb]{$\mu$}
\Text(25,8)[cb]{${\scriptstyle W, G_W}$}
\Text(45,-22)[cb]{${\scriptstyle W}$}
\Text(60,24)[cb]{${\scriptstyle W}$}
\Text(15,30)[cb]{$b$}
\end{picture}
&\quad
\begin{picture}(100,50)(0,0)
\Photon(0,0)(30,0){1.3}{6}
\Line(90,34.5)(70,23)
\Photon(70,23)(50,11.5){1.3}{5}
\Line(50,11.5)(30,0)
\Line(90,-34.5)(70,-23)
\Photon(70,-23)(50,-11.5){1.3}{5}
\Line(50,-11.5)(30,0)
\Line(50,-11.5)(50,11.5)
\Line(70,-23)(70,23)
\Text(5,-12)[cb]{$\mu$}
\Text(46,-27)[cb]{$\scriptstyle \gamma,Z,W$}
\Text(46,19)[cb]{$\scriptstyle \gamma,Z,W$}
\Text(15,30)[cb]{$c$}
\end{picture}
\end{tabular}
\vspace{1.5cm}
}
\caption[]{Diagrams contributing to the fermionic corrections to 
$\sin^2\theta_{\rm eff}^{\rm lept}$.
Because of CP conservation, diagrams involving the Higgs boson 
cancel and are not included.}
\label{sin2}
\end{figure}
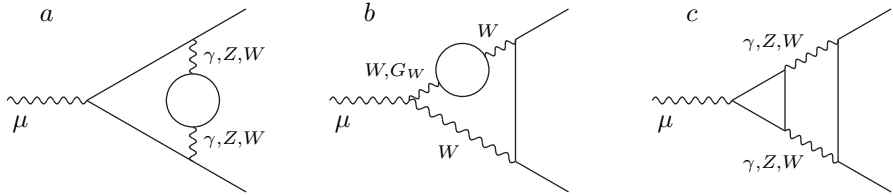

The effective leptonic weak mixing angle is at present the observable
that can give the most stringent indirect evaluation on the 
mass of the Higgs boson (by means of the radiative corrections).
In the future, if the Higgs boson would be discovered, 
$\sin^2\theta_{\rm eff}^{\rm lept}$ would represent 
a strong test of the Standard Model.
It is defined throught the vectorial and axial effective couplings $g_v$ and 
$g_a$ of the Z boson with leptons:
\bq
\sin^2\theta_{eff}^l= 
\frac{1}{4}\,\left( 1 - {\rm Re}\,\left(\frac{g_v}{g_a}\right) \right).
\eq
The effective couplings $g_v$ and $g_a$ are defined throught the amplitude 
of the decay of an on-shell Z boson into two leptons:
\bq
{\cal M}^l_{Zll}= 
{\bar u}_l\,{\cal M}^{\mu}\,v_l\,\epsilon_{_Z}^{\mu}= 
{\bar u}_l\,\gamma_{\mu}\,(g_v + g_a\,\gamma_5 )\,v_l\,\epsilon_{_Z}^{\mu},
\qquad\quad
s= M_Z^2.
\eq
Therefore $g_v$ and $g_a$ can be obtained from the matrix ${\cal M}^{\mu}$ 
by using suitable projectors:
\bq
\frac{1}{D}\,{\rm Tr}\,(\gamma_{\mu}\,{\cal M}^{\mu}),
\qquad
g_{a} = 
-\,\frac{1}{D}\,{\rm Tr}\,(\gamma_{\mu}\,\gamma_5\,{\cal M}^{\mu}),
\eq
where $D$ is the dimension of the space-time.
Among the electroweak two-loop diagrams contributing to ${\cal M}^{\mu}$, 
we started considering those containing a closed fermion loop 
(this computation has been recently done by M.~Awramik, M.~Czakon, 
A.~Freitas and G.~Weiglein in~\cite{Awramik:2004ge}).
They are represented in \fig{sin2}.

We started appling our methods on configuration $c$ which is the most 
difficult one.
The strategy used is the following:
\bei
\item
Write the amplitude, considering the different contributions to 
configuration $c$.
This generates a sum of tensor integral of the family $V^{231}$.
\item
Perform a simple reduction of tensor integral of the following type:
\bqa
&{}&\qquad\qquad\qquad
\frac{2\,\spro{q}{p}}{(q^2+m^2)\,[(q+p)^2+M^2]}=
\nl
&{}&
\frac{1}{q^2+m^2} \,-\, \frac{1}{(q+p)^2+M^2}
\,-\,\frac{p^2-m^2+M^2}{(q^2+m^2)\,[(q+p)^2+M^2]}.
\eqa
This kind of reduction does not introduce any new denominator and 
therefore any spurious singularity.
At this point we have a sum of scalar and vector integrals belonging to 
4 vertex families ($V^{121}$, $V^{131}$, $V^{221}$ and $V^{231}$), 
together with some self-energies and some one-loop diagrams.
\item
Combine the sum of all these diagrams in just {\itshape one} integral.
\eei
Just to show the efficiency of the numerical computation, we give 
some preliminary results.
We consider the expansion in loops for $g_v/g_a$
\bq
g_{v,a}= g_{v,a}^0 + g_{v,a}^1 + g_{v,a}^2,
\eq
\bq
\frac{g_v}{g_a}= 
\frac{g_v^0}{g_a^0}\,
\left[
  1
+ \frac{g_v^1}{g_v^0} - \frac{g_a^1}{g_a^0}
- \frac{g_a^1}{g_a^0}\,
  \left( \frac{g_v^1}{g_v^0} - \frac{g_a^1}{g_a^0} \right)
+ \frac{g_v^2}{g_v^0} - \frac{g_a^2}{g_a^0}
\right],
\eq
where the last two terms represent the pure two-loop corrections to 
$\sin^2\theta_{\rm eff}^{\rm lept}$.
The contribution to these corrections coming from diagram $c$ (\fig{sin2})
with two $Z$ or two $W$ is:
\bqa
\left( \frac{g_v^2}{g_v^0} - \frac{g_a^2}{g_a^0} \right)_{ZZ}
&=&
- 0.279937 \times 10^{-5} \,\pm\, 0.15 \times 10^{-9},
\nl
\left( \frac{g_v^2}{g_v^0} - \frac{g_a^2}{g_a^0} \right)_{WW}
&=&
0.577269 \times 10^{-1} \,\pm\, 0.14 \times 10^{-5}.
\eqa
In this result are summed up the contributions coming from 
all possible fermion loops.
In particular it includes the diagrams containing the top quark which
in the usual analytical approach require an expansion in the ratio 
$M_Z/m_t\sim 1/4$.
\section{Conclusions}
The techniques described in this paper show that the numerical approach 
to Feynman integrals allows the computation of diagrams that cannot be 
treated within the usual analytical approach.
Under this point of view, would be interesting to apply these methods, 
and in particular the new one, to more complicated two-loop diagrams 
(the two-loop 4-point functions for example).
In addition to that, the first results, obtained from the application 
of these techniques to the fermionic correction to 
$\sin^2\theta_{\rm eff}^{\rm lept}$, seem to show that the 
numerical approach is not only suited for the computation of single 
integrals, but can also be applied to the complete evaluation of 
physical observables (which requires to sum up several diagrams).
Of course the computaion must be completed to give a serious proof 
of that.
\section*{Acknowledgments}
I would like to thank A.~Ferroglia, M.~Passera and G.~Passarino for the 
collaboration on the computation of two-loop vertices and W.~Hollik and 
U.~Meier for the common work on the evaluation of $\sin^2_{eff}$. 
\appendix
\section{}
\begin{figure}[ht]
{\small
\noindent
\begin{tabular}{ccc}
\begin{picture}(100,50)(0,0)
\Line(0,0)(30,0)
\Line(90,34.5)(70,23)
\CArc(70,-23)(46,90,150)
\CArc(30,46)(46,-90,-30)
\Line(90,-34.5)(30,0)
\Line(70,-23)(70,23)
\Text(10,4)[cb]{$-P$}
\Text(86,-44)[cb]{$p_1$}
\Text(86,38)[cb]{$p_2$}
\Text(40,40)[cb]{$V^{121}$}
\end{picture}
&\quad
\begin{picture}(100,50)(0,0)
\Line(0,0)(30,0)
\Line(90,34.5)(47,10)
\CArc(39,6)(10,0,360)
\Line(90,-34.5)(30,0)
\Line(70,-23)(70,23)
\Text(10,4)[cb]{$-P$}
\Text(86,-44)[cb]{$p_1$}
\Text(86,38)[cb]{$p_2$}
\Text(32,18)[cb]{$m_1$}
\Text(57,-4)[cb]{$m_2$}
\Text(56,21)[cb]{$m_3$}
\Text(80,-4)[cb]{$m_4$}
\Text(50,-22)[cb]{$m_5$}
\Text(40,40)[cb]{$V^{131}$}
\end{picture}
&\quad
\begin{picture}(100,50)(0,0)
\Line(0,0)(70,0)
\Line(90,34.5)(30,0)
\Line(90,-34.5)(30,0)
\Line(70,-23)(70,23)
\Text(10,4)[cb]{$-P$}
\Text(86,-44)[cb]{$p_1$}
\Text(86,38)[cb]{$p_2$}
\Text(40,40)[cb]{$V^{221}$}
\end{picture}
\end{tabular}
\vspace{2cm}

\noindent
\begin{tabular}{ccc}
\begin{picture}(100,50)(0,0)
\Line(0,0)(30,0)
\Line(90,34.5)(58.5,16.3)
\Line(41,7)(30,0)
\CArc(50,11.5)(10,0,360)
\Line(90,-34.5)(30,0)
\Line(70,-23)(70,23)
\Text(10,4)[cb]{$-P$}
\Text(86,-44)[cb]{$p_1$}
\Text(86,38)[cb]{$p_2$}
\Text(40,40)[cb]{$V^{141}$}
\end{picture}
&\quad
\begin{picture}(100,50)(0,0)
\Line(0,0)(30,0)
\Line(90,34.5)(30,0)
\Line(90,-34.5)(30,0)
\Line(50,-11.5)(50,11.5)
\Line(70,-23)(70,23)
\Text(10,4)[cb]{$-P$}
\Text(86,-44)[cb]{$p_1$}
\Text(86,38)[cb]{$p_2$}
\Text(35,-15)[cb]{$m_1$}
\Text(35,9)[cb]{$m_2$}
\Text(59,-4)[cb]{$m_3$}
\Text(56,-26)[cb]{$m_4$}
\Text(80,-4)[cb]{$m_5$}
\Text(56,21)[cb]{$m_6$}
\Text(40,40)[cb]{$V^{231}$}
\end{picture}
&\quad
\begin{picture}(100,50)(0,0)
\Line(0,0)(30,0)
\Line(90,34.5)(30,0)
\Line(90,-34.5)(30,0)
\Line(50,-11.5)(70,23)
\Line(70,-23)(57.7,-2)
\Line(55.5,2)(50,11.5)
\Text(10,4)[cb]{$-P$}
\Text(86,-44)[cb]{$p_1$}
\Text(86,38)[cb]{$p_2$}
\Text(40,40)[cb]{$V^{222}$}
\end{picture}
\end{tabular}
\vspace{1.8cm}
}
\caption[]{Two-loop vertex topologies. The momenta are all incoming.}
\label{vertices}
\end{figure}
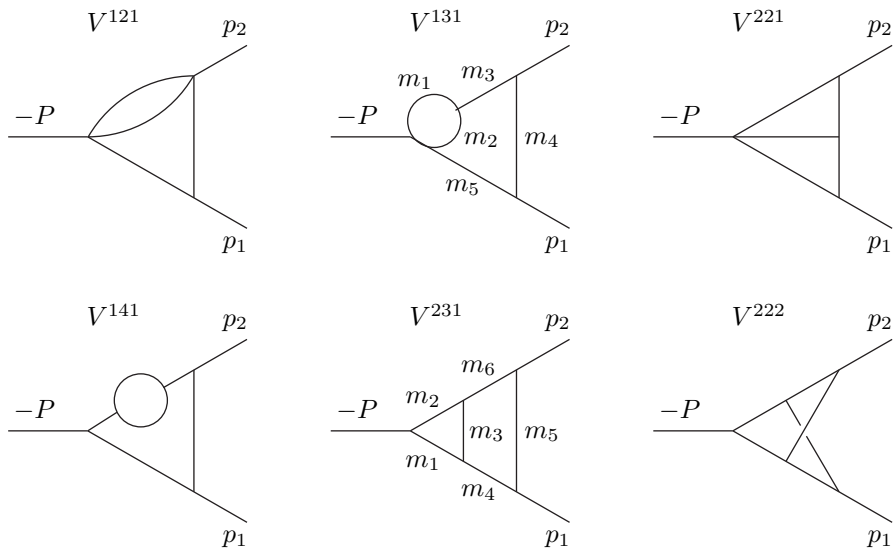


\begin{thebibliography}{99}

\bibitem{unknown:2003ih}
  [LEP Collaboration],
arXiv:hep-ex/0312023.

\bibitem{Davydychev:2002}
A.I.~Davydychev and V.A.~Smirnov,
Nucl.\ .Instrum.\ Meth.\ A {\bf 502} (2003) 621.
[hep-ph/0210171].

\bibitem{Remiddi:2003}
R.~Bonciani, P.~Mastrolia and E.~Remiddi,
Nucl.\ Phys.\ B {\bf 661} (2003) 289.
[hep-ph/0301170].

\bibitem{Tkachov:1997wh}
F.~V.~Tkachov,
Nucl.\ Instrum.\ Meth.\ A {\bf 389} (1997) 309
[hep-ph/9609429];\\
L.~N.~Bertstein, Functional Analysis and its Applications, {\bf 6}(1972)66.

\bibitem{Passarino:2001wv}
G.~Passarino,
Nucl.\ Phys.\ B {\bf 619} (2001) 257
[arXiv:hep-ph/0108252].

\bibitem{Passarino:2001jd}
G.~Passarino and S.~Uccirati,
Nucl.\ Phys.\ B {\bf 629} (2002) 97
[arXiv:hep-ph/0112004].

\bibitem{Ferroglia:2002mz}
A.~Ferroglia, M.~Passera, G.~Passarino and S.~Uccirati,
Nucl.\ Phys.\ B {\bf 650} (2003) 162
[arXiv:hep-ph/0209219].

\bibitem{Ferroglia:2002yr}
A.~Ferroglia, G.~Passarino, M.~Passera and S.~Uccirati,
{\it Prepared for 31st International Conference on 
High Energy Physics 
(ICHEP 2002), Amsterdam, The Netherlans};\\
A.~Ferroglia, G.~Passarino, S.~Uccirati and M.~Passera,
Nucl.\ Instrum.\ Meth.\ A {\bf 502} (2003) 391.

\bibitem{Ferroglia:2003yj}
A.~Ferroglia, M.~Passera, G.~Passarino and S.~Uccirati,
Nucl.\ Phys.\ B {\bf 680} (2004) 199
[arXiv:hep-ph/0311186].

\bibitem{Actis:2004bp}
S.~Actis, A.~Ferroglia, G.~Passarino, M.~Passera and S.~Uccirati,
arXiv:hep-ph/0402132.

\bibitem{ellip} R.~Erdelyi et al., 
{\it Higher Transcendental Functions 
vol. 2}, Bateman Manuscript Project ({\it McGraw-Hill 1953});\\
L.~J.~Slater, {\it Generalized Hypergeometric Functions}
Cambridge Univ. Press, 1966.

\bibitem{Awramik:2004ge}
M.~Awramik, M.~Czakon, A.~Freitas and G.~Weiglein,
arXiv:hep-ph/0407317.

\end{thebibliography}
\end{document}